\documentclass[manuscript]{aastex631}
\usepackage{amsmath}
\usepackage{graphics}
\usepackage{natbib}
\usepackage{amsmath}
\usepackage{amssymb}
\usepackage{longtable,booktabs}
\usepackage{array, makecell}

\begin{document}
\title{Inertial-range magnetic fluctuation anisotropy observed from PSP's first seven orbits}
\author{L.-L. Zhao}
\affiliation{Center for Space Plasma and Aeronomic Research (CSPAR), The University of Alabama in Huntsville, Huntsville, AL 35805, USA}
\affiliation{Department of Space Science, The University of Alabama in Huntsville, Huntsville, AL 35805, USA}
\author{G. P. Zank}
\affiliation{Center for Space Plasma and Aeronomic Research (CSPAR), The University of Alabama in Huntsville, Huntsville, AL 35805, USA}
\affiliation{Department of Space Science, The University of Alabama in Huntsville, Huntsville, AL 35805, USA}
\author{L. Adhikari}
\affiliation{Center for Space Plasma and Aeronomic Research (CSPAR), The University of Alabama in Huntsville, Huntsville, AL 35805, USA}
\affiliation{Department of Space Science, The University of Alabama in Huntsville, Huntsville, AL 35805, USA}
\author{M. Nakanotani}
\affiliation{Center for Space Plasma and Aeronomic Research (CSPAR), The University of Alabama in Huntsville, Huntsville, AL 35805, USA}
\affiliation{Department of Space Science, The University of Alabama in Huntsville, Huntsville, AL 35805, USA}

\begin{abstract}
Solar wind turbulence is anisotropic with respect to the mean magnetic field. Anisotropy leads to ambiguity when interpreting in-situ turbulence observations in the solar wind because an apparent change in the measurements could be due either to the change of intrinsic turbulence properties or to a simple change of the spacecraft sampling direction.
We demonstrate the ambiguity using the spectral index and magnetic compressibility in the inertial range observed by the \textit{Parker Solar Probe} during its first seven orbits ranging from 0.1 to 0.6 AU.
	To unravel the effects of the sampling direction, we assess whether the wavevector anisotropy is consistent with a two-dimensional (2D) plus slab turbulence transport model and determine the fraction of power in the 2D versus slab component.
Our results confirm that the 2D plus slab model is consistent with the data and the power ratio between 2D and slab components depends on radial distance, with the relative power in 2D fluctuations becoming smaller closer to the Sun.
\end{abstract}

\section{Introduction}\label{sec:introduction}

The nature of anisotropy is an important property of solar wind turbulence.
Decades of in-situ solar wind observations have found that low-frequency turbulence is dominated by incompressible fluctuations. The magnetic fluctuations are mostly transverse to the mean magnetic field.
Solar wind turbulence can exhibit wavevector anisotropy in that more power is often found in fluctuations with perpendicular wavevectors.
The nearly incompressible theory of MHD turbulence suggests that solar wind turbulence in the plasma $\beta \sim 1$ or $\ll$ 1 regimes is dominated by a 2D component whose wavevectors are perpendicular to the mean magnetic field with a minority contribution due to slab fluctuations whose wavevectors are along the mean magnetic field \citep{matthaeus1990, Zank1992JGR, Zank1993PF, Hunana2010, Zank2017ApJ}.
The 2D and slab turbulence model is consistent with many previous observations \citep{Bieber1994, bieber1996, zank1996evolution, smith2001, oughton2011, wiengarten2016, Adhikari2017APJ, adhikari2020solar}, and can explain the observed wavevector anisotropy of solar wind turbulence, i.e., the fluctuation power tends to be stronger when the spacecraft samples perpendicular to the mean magnetic field \citep{Zank2020ApJ}.
The compressible fluctuations in the nearly incompressible (NI) theory arise at the second order, and the level of compressibility depends on parameters such as the plasma beta, but the exact relation is unclear.

The wavevector anisotropy of solar wind turbulence can be assessed based on the sampling angle between the spacecraft trajectory and the mean magnetic field thanks to Taylor's hypothesis \citep{horbury2008, dasso2005, wang2019}. Since the solar wind velocity is usually in the radial direction, the sampling angle can be approximated by the angle between mean magnetic field and radial direction.
By compiling measurements of solar wind turbulence with angular dependence, the wavevector anisotropy can be investigated.
Such studies have been undertaken by numerous authors. For example, \cite{matthaeus1990} investigated the anisotropic turbulence correlation function, which suggests an 2D and slab decomposition of turbulence. \cite{bieber1996} developed a method that quantifies the power fraction in the 2D and slab components based on the functional form of the angular dependence. Their results suggest that $\sim$ 80\% of the energy is contained in the 2D turbulence, which is consistent with the turbulence anisotropy inferred from cosmic ray mean free path observations \citep{zhao2017cosmic, zhao2018influence}.
A particular interesting situation is where the wavevector is purely parallel, which has been studied recently by \cite{telloni2019no, zhao2020spectral}. Both reported a $-5/3$ power law index, which is inconsistent with the ``critical balance'' prediction of -2 \citep{goldreich1995} but can be explained by NI MHD theory \citep{Zank2020ApJ, zank2021turbulence}.
The wavelet technique has also been applied to the study of wavevector anisotropy \citep[e.g.,][]{horbury2008}. The wavelet method considers a scale-dependent local mean magnetic field, which may differ from the global mean field especially when the fluctuations are strong.
We opt not to use the wavelet method as the 2D/slab turbulence geometry requires a strong global mean field to sensibly distinguish between the two components \citep{oughton2020}.

The Parker Solar Probe (PSP) \citep{fox2016solar} provides in-situ measurements of solar wind turbulence within 0.3 AU from the Sun for the first time. In this paper, we investigate magnetic turbulence anisotropy in the inertial range using PSP data, which allows us to assess the radial evolution in the inner heliosphere.
Previous work by \cite{chen2020} considered the magnetic compressibility as defined by the ratio between magnetic field magnitude fluctuations and total magnetic field fluctuations. Using the PSP data from the first orbit, they conclude that magnetic compressibility increases with radial distance and plasma beta, which they interpret as a change in the slow magnetosonic mode fraction with radial distance. However, the angular dependence or wavevector anisotropy is not taken into account by \cite{chen2020}.
As the interplanetary magnetic field follows the Parker spiral shape \citep{parker1958} on average, the angle it makes with radial direction also changes with radial distance from the Sun, which then affects the turbulence properties. This effect needs to be disentangled from observations before the nature of the radial evolution of turbulence can be revealed.
This effect has not been considered in previous work using PSP data. We present an analysis of turbulence anisotropy based on its angular dependence. We further quantify the wavevector anisotropy based on the methodology of \cite{bieber1996} and illustrate its radial evolution.

\section{Methods}\label{sec:m}
One way to quantify the anisotropy is to calculate several indices based on the 2D/slab turbulence model. Here, we use the method developed by \cite{bieber1996}.
The magnetic field fluctuations are transformed into coordinates where the $z$ axis represents the mean magnetic field direction, the $x$ axis is in the plane containing the radial direction and $z$ direction, and the $y$ axis completes a right-handed system, which is perpendicular to both radial and $z$ direction.
The power spectral density (PSD) can then be calculated in the new coordinates. We note that the spacecraft measured PSD used in subsequent analysis is the 1D reduced spectrum. The diagonal components of the spectral matrix $P_{xx}$, $P_{yy}$, and $P_{zz}$ are the power contained in the fluctuations along the three coordinates. Evidently, $P_{zz}$ represents the longitudinal fluctuation power, and $P_{xx}$ and $P_{yy}$ denote the incompressible transverse fluctuations. The ratio $(P_{xx} + P_{yy}) / P_{zz}$ can be used as a measure of the compressibility, i.e., the higher the ratio, the less compressible is the turbulence. This stems from the limit of small amplitude MHD waves where $P_{zz}$ is nonzero only for compressible magnetosonic waves. However, this is not valid for large-amplitude (aka spherically polarized) Alfv\'en waves, which can have a finite $P_{zz}$ \citep{Barnes1974, Matteini2015}.
For the incompressible part of the turbulence, when the mean field is along the radial direction, the spacecraft will sample mostly slab fluctuations by Taylor's hypothesis. In this case, one would expect $P_{xx} = P_{yy}$ if there is no anisotropy in the perpendicular plane. As the angle between the mean magnetic field and radial direction $\theta_{B R}$ increases, the spacecraft can sample more 2D fluctuations with perpendicular wavevectors. At $\theta_{B R} = 90^{\circ}$, only 2D fluctuations are sampled with measurably different $P_{xx}$ and $P_{yy}$. The change of the ratio $P_{yy} / P_{xx}$ with $\theta_{B R}$ is thus a measure of the 2D versus slab power ratio.
More precisely, we use the formula derived in \cite{bieber1996} for the ratio between the 2D and slab components \citep[see also][]{saur1999},
\begin{equation}\label{eq:Pyyxx}
  \frac{P_{yy}}{P_{xx}} = \frac{(1+q) C_s \cos^{q-1} \theta_{B R} + 2q C_2 \sin^{q-1} \theta_{B R}}{(1+q)C_s \cos^{q-1} \theta_{B R} + 2C_2 \sin^{q-1} \theta_{B R}}.
\end{equation}
Here, $C_2$ and $C_s$ represent the relative amplitude of power in 2D and slab components, and $\theta_{B R}$ the angle between the mean magnetic field $\boldsymbol{B}_0$ and the radial direction. In deriving Equation \eqref{eq:Pyyxx}, it has been assumed that the fluctuations follow power-law distributions in which the spectral indices $q$ are the same for both perpendicular ($P_{yy}$) and parallel ($P_{xx}$) fluctuations.
Equation \eqref{eq:Pyyxx} shows that for pure slab turbulence ($C_2 = 0$), the ratio is unity for all sampling angles $\theta_{B R}$. In the other limit of $C_s = 0$, the ratio is also constant for all $\theta_{B R}$, but has a different value depending on $q$.
In general, Equation \eqref{eq:Pyyxx} allows us to infer $C_s$ and $C_2$ by probing the variation of the ratio $P_{yy}/P_{xx}$ as a function of the angle $\theta_{B R}$. This is the basis of our subsequent analysis. \cite{Zank2020ApJ} have generalized expression \eqref{eq:Pyyxx} to allow for different spectral indices $q_{yy}$ and $q_{xx}$ for the 2D and slab power law distributions. This more extended analysis will be presented in a subsequent paper.

\section{\textit{PSP} Data Overview and Results}\label{sec:results}
We use Level 2 magnetic field data measured by \textit{PSP} FIELD/MAG instrument \citep{Bale2016} and Level 3 plasma proton data measured by the SWEAP/SPC instrument \citep{Kasper2016}. Our analysis uses the first seven orbit measurements of \textit{PSP} during the period from 2018 November to 2021 February. However, the third orbit is excluded due to a large number of data gaps in magnetic field measurements and SPC plasma data are not available during the third outbound trajectory \citep{zhao2021detection}. For each orbit, we investigate an about one-month time period data around each perihelion, with radial distances ranging from each perihelion to 0.6 AU. The magnetic fluctuation anisotropy of each orbit is analyzed separately, and the time intervals for analysis are chosen to be 1-hour in length. It is well known that various solar wind turbulence features are speed-related \citep[e.g.,][]{dasso2005, weygand2011, adhikari2021c}, and the fast wind is usually found to be dominated by outward propagating Alfv\'en waves. Here, we exclude intervals with an average speed greater than 500 km/s when plasma data are available. To remove intervals associated with large-scale structures, e.g., heliospheric/strong current sheet crossings \citep{phan2021}, two criteria are used in each interval: (\romannumeral 1) the standard deviation of the angle between local magnetic field at each data point and the radial direction should be smaller than 40 degrees, and (\romannumeral 2) the ratio between the variance in the magnetic field magnitude fluctuations and the variance in the total magnetic field fluctuations should be smaller than 0.3. If any of these criteria are not met, the interval is discarded. For the remaining intervals, we calculate the angle between the mean magnetic field (estimated in 1-hour interval) and the radial direction $\theta_{BR}$. We require that $0^{\circ} \leq \theta_{BR} \leq 90^{\circ}$ for each interval and do not distinguish between sunward and anti-sunward directions. The in-situ measured magnetic field data in RTN coordinates with a cadence of $\sim$0.22 s is then projected to the mean field-aligned $XYZ$ coordinate system defined in Section \ref{sec:m}. After projection, we calculate the PSD of each component (i.e., $B_X$, $B_Y$, and $B_Z$) using a standard Fourier transform method. To describe the inertial range fluctuations, the PSD of each component and the total trace spectrum in the frequency range of 0.01 Hz $\leq f_{sc} \leq$ 0.1 Hz is fitted to a power law separately when a power law form is clearly observed. We have determined that the selected frequency range falls well within the inertial range for the radial distance at 0.1--0.6 AU. The fitted spectral exponents and amplitudes of the fluctuations calculated by integrating the PSD over the frequency range 0.01 Hz -- 0.1 Hz are used in the following anisotropy analysis. 

\subsection{Radial and sampling angle dependence of turbulence}
\begin{figure}[htbp]
\centering
\includegraphics[width=0.5\linewidth]{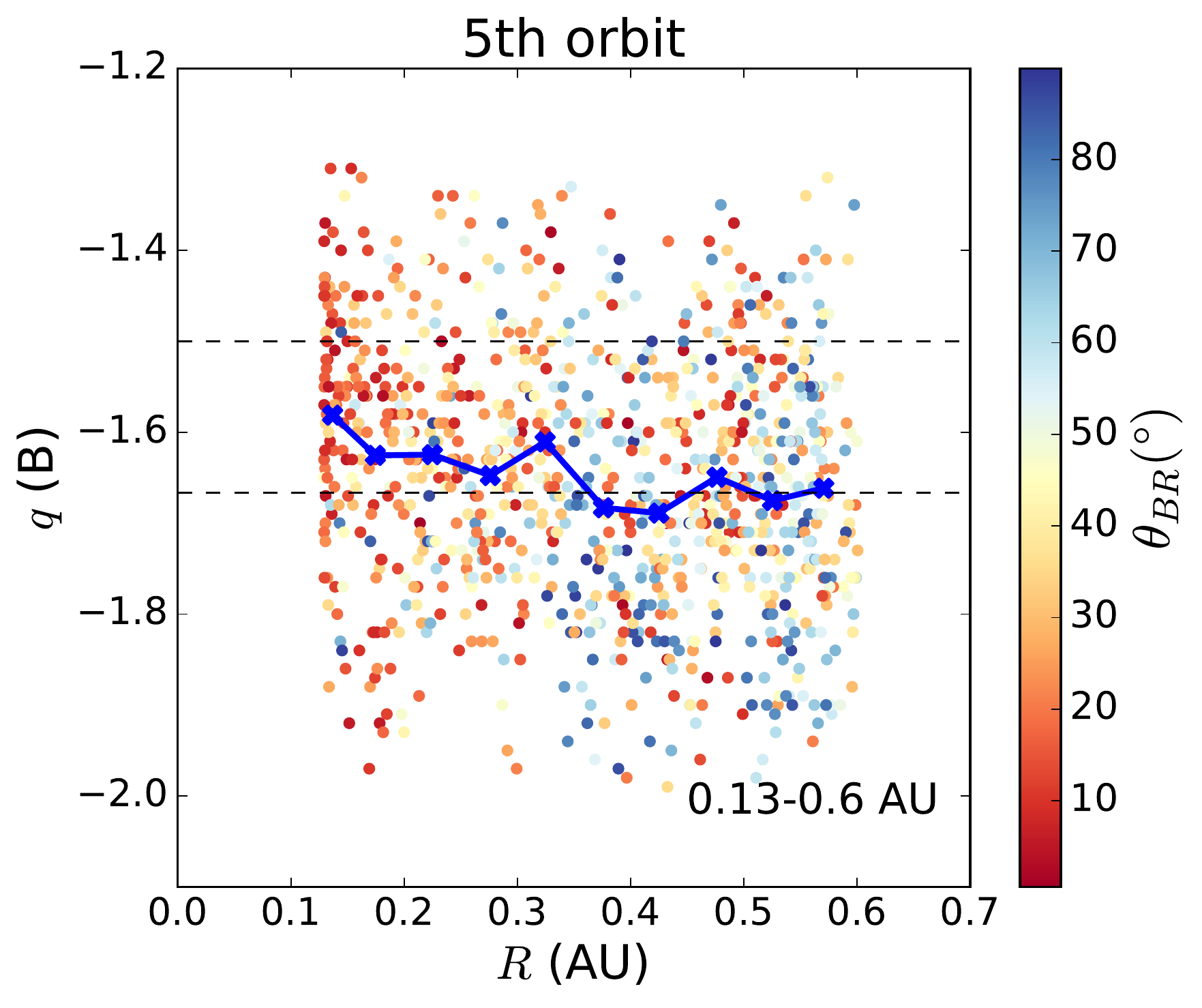}%
\includegraphics[width=0.5\linewidth]{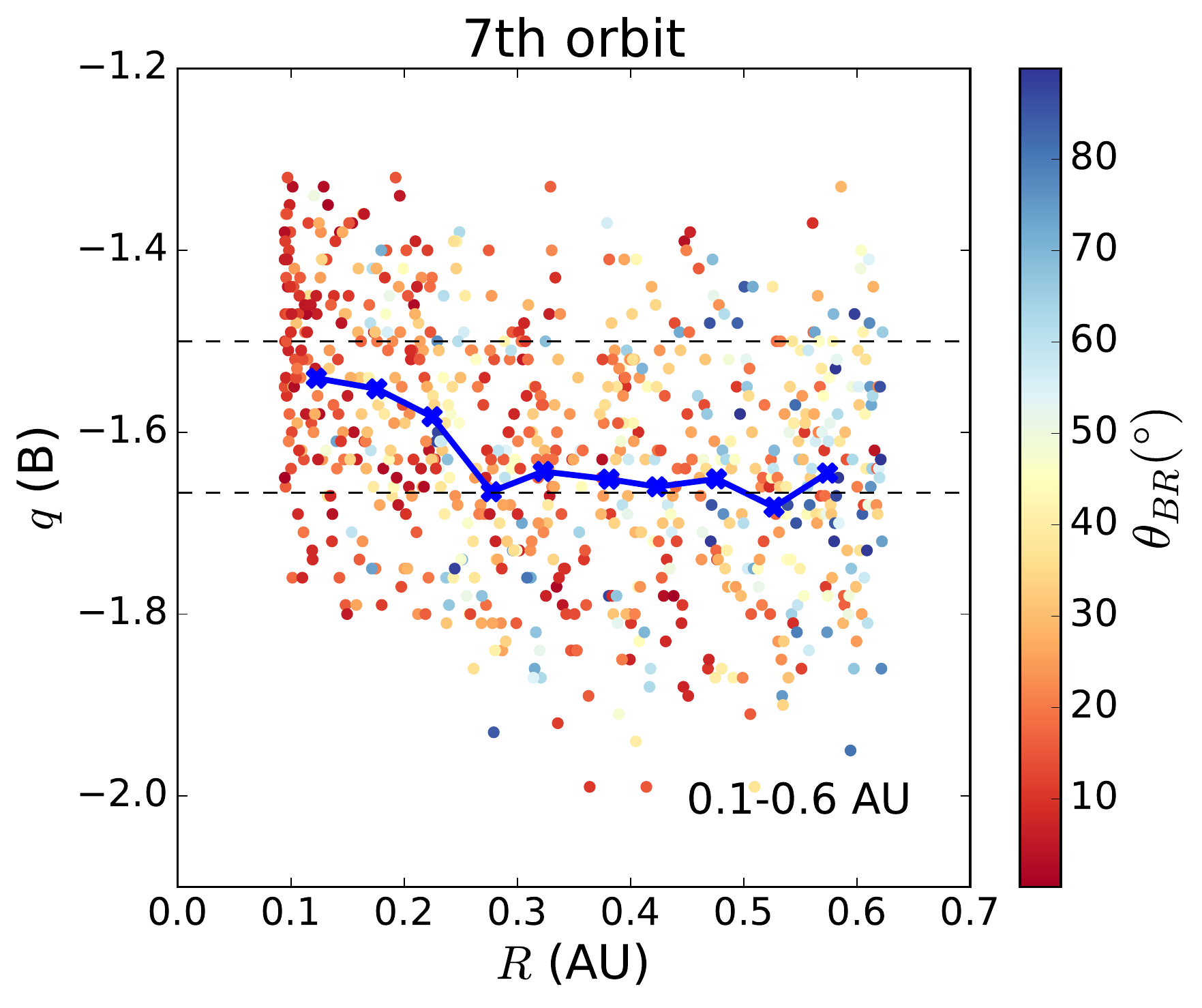}
	\caption{The power-law index $q$ of the magnetic trace PSD as a function of radial distance $R$ during \textit{PSP}'s fifth and seventh orbits. The color map indicates the angle between the mean magnetic field in each 1-hour interval and the radial direction $\theta_{BR}$. The blue lines with crosses represent the average values of the spectral indices.} \label{fig:view1103}
\end{figure}

We analyze the magnetic field data from the first seven orbits of the \textit{PSP}, and we present mainly the results of the fifth orbit here as examples.
Figure \ref{fig:view1103} shows the power-law index $q$ of the magnetic trace spectrum over the frequency range of 0.01 Hz $\leq f_{sc} \leq$ 0.1 Hz as a function of radial distance $R$ and color coded by angle $\theta_{BR}$ during \textit{PSP}'s fifth (left panel) and seventh (right panel) orbits. The data are from orbit 5 between 2020 May 15 and June 28, with a radial distance of 0.13 AU to 0.6 AU. The seventh orbit data are from 2021 January 6 to February 9 and the radial distance ranges from 0.1 to 0.6 AU. Two horizontal dashed lines indicate the $-1.5$ Iroshnikov-Kriachnan (IK) spectrum and $-5/3$ Kolmogorov spectrum. The blue line with crosses shown in each panel represents the average values of the spectral indices, which are calculated in $0.05$ AU bins.

The figure suggests that the spectrum is shallower (smaller $|q|$) closer to the Sun, which is consistent with previous results \citep{chen2020}. However, the color map indicates that the radial dependency may depend quite strongly on the angle $\theta_{BR}$ which in turn tends to increase with increasing radial distance.
It is unclear whether the variation in power-law index is due to the radial evolution of intrinsic turbulence properties or the sampling direction of the spacecraft.

\begin{figure}[htbp]
\centering
\includegraphics[width=0.5\linewidth]{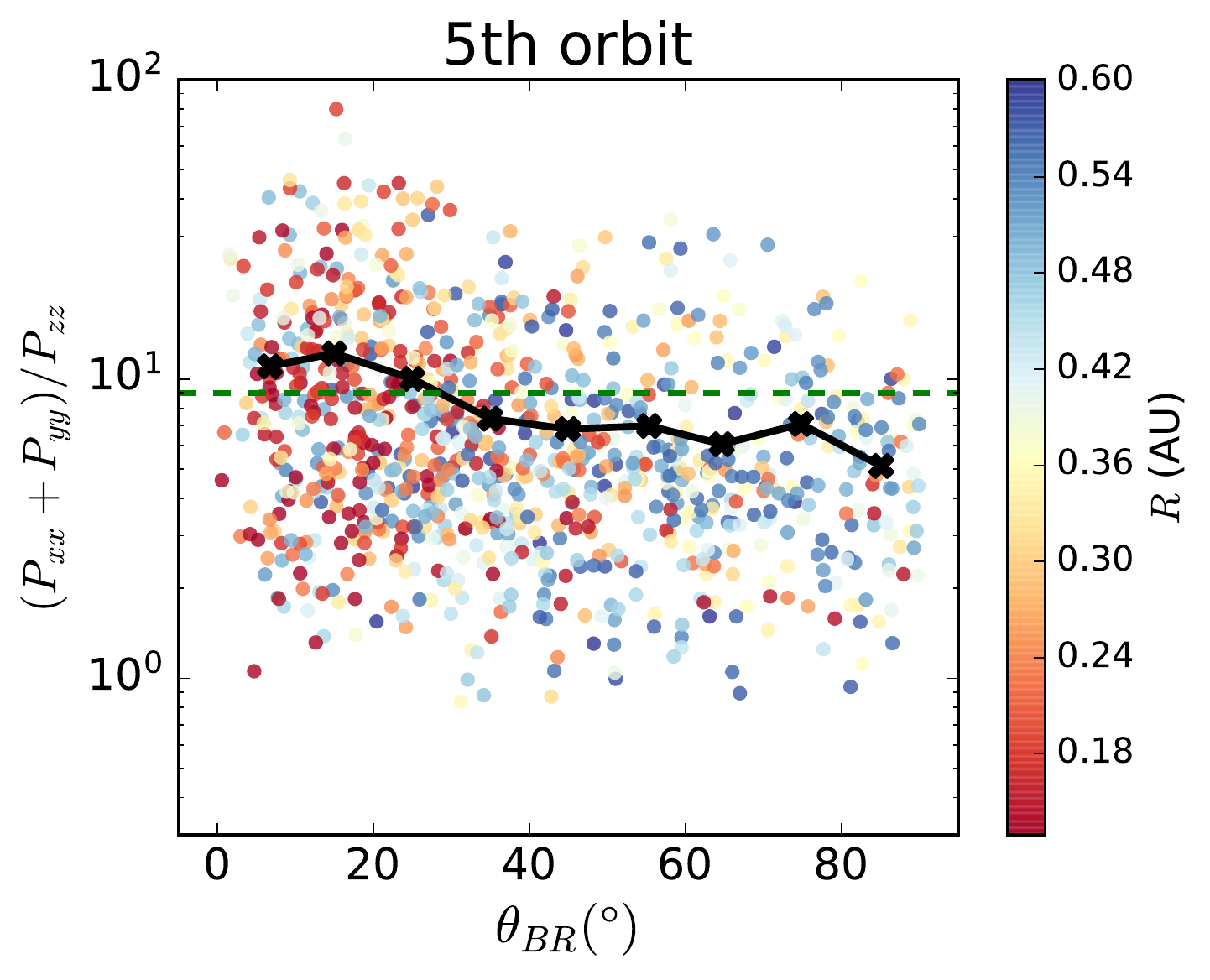}%
\includegraphics[width=0.5\linewidth]{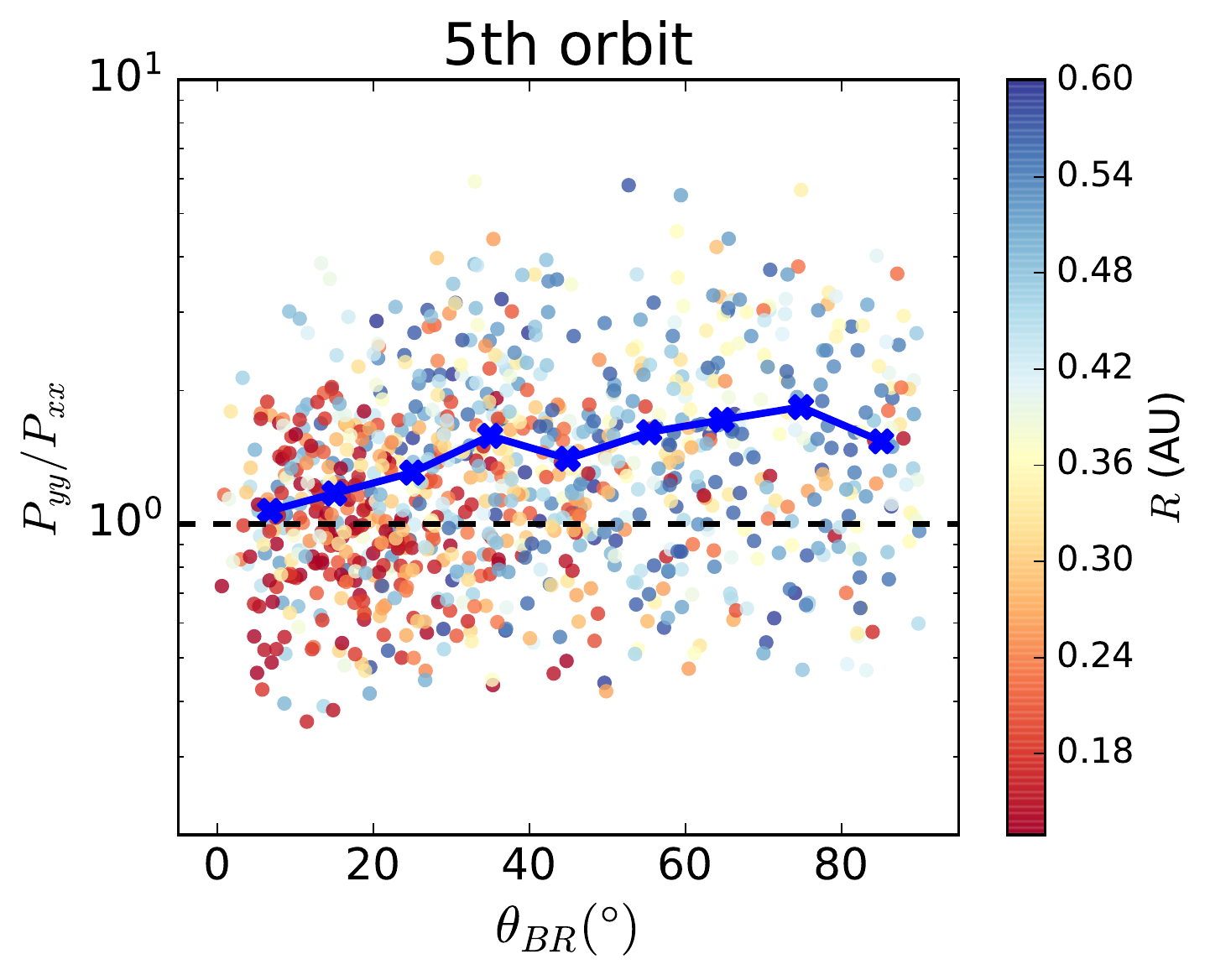}
	\caption{The left panel shows statistics of the transverse-to-longitudinal $(P_{xx}+P_{yy})/P_{zz}$ power anisotropy (compressibility) as a function of $\theta_{BR}$ during \textit{PSP}'s fifth orbit. The right panel shows the transverse power anisotropy $P_{yy}/P_{xx}$ as a function of $\theta_{BR}$. The color map indicates the radial distance $R$ dependence. The black and blue lines with crosses in left and right panels represent the average values of $(P_{xx}+P_{yy})/P_{zz}$ and $P_{yy}/P_{xx}$, respectively.}\label{fig:psd1}
\end{figure}

In the left panel of Figure \ref{fig:psd1}, we show the dependence of the transverse-to-longitudinal power anisotropy $(P_{xx}+P_{yy})/P_{zz}$, which is a measure of the magnetic compressibility, on the sampling direction $\theta_{BR}$ and radial distance $R$. The horizontal dashed line corresponds to the well-known value of $(P_{xx}+P_{yy})/P_{zz} = 9$ found by \cite{belcher1971large} originally using Mariner 5 data. The black line with crosses denotes the average value of $(P_{xx}+P_{yy})/P_{zz}$ computed over nine bins and each bin has a width of 10$^\circ$. 

When \textit{PSP} is close to the Sun, the sampling angle $\theta_{BR}$ tends to be quasi-parallel. In general, the power anisotropy $(P_{xx}+P_{yy})/P_{zz}$ decreases as does the sampling angle $\theta_{BR}$ increases. When $\theta_{BR}$ is quasi-parallel, the average value of $(P_{xx}+P_{yy})/P_{zz}$ is larger than 9. However, when $\theta_{BR}$ increases to greater than $30^\circ$, $(P_{xx}+P_{yy})/P_{zz}$ is obviously less than 9 and the measured fluctuations appear to be more compressible when $\theta_{BR}$ approaches $90$ degrees. Similar to Figure \ref{fig:view1103}, the color map indicates that the angular dependence and the radial dependence may be related.
The right panel of Figure \ref{fig:psd1} shows the transverse fluctuation power anisotropy $P_{yy}/P_{xx}$ as a function of the field orientation $\theta_{BR}$ and color coded by the radial distance $R$. The blue line with crosses represents the average value of $P_{yy}/P_{xx}$ in each 10$^\circ$ bin. It's clear that $P_{yy}/P_{xx}$ increases as $\theta_{BR}$ approaches 90$^\circ$, which tends to occur at larger distances. The changes of $P_{yy}/P_{xx}$ with $\theta_{BR}$ is generally consistent with Equation \eqref{eq:Pyyxx} and implies again that the field orientation is important and cannot be neglected in determining the overall power anisotropy. We note that both of $P_{yy}$ and $P_{xx}$ depend not only on $\theta_{BR}$ but also on the radial distance $R$. This is partly because the $\theta_{BR}$ and $R$ are themselves related, i.e., $\theta_{BR}$ tends to increase with $R$. However, there may be transport effects associated with radial distance $R$ as well, e.g, $P_{xx}(R) \propto R^\alpha$, $P_{yy}(R) \propto R^\beta$, and $\alpha \ne \beta$. In fact, all three components, $P_{xx}$, $P_{yy}$, and $P_{zz}$ can change differently with $\theta_{BR}$ at a fixed radial distance, possibly including the effects of sampling. In Section \ref{sec:2Dslab}, we attempt to isolate the angular dependence of $P_{yy}/P_{xx}$ from the radial dependence. 

\begin{figure}[htbp]
\centering
\includegraphics[width=0.5\linewidth]{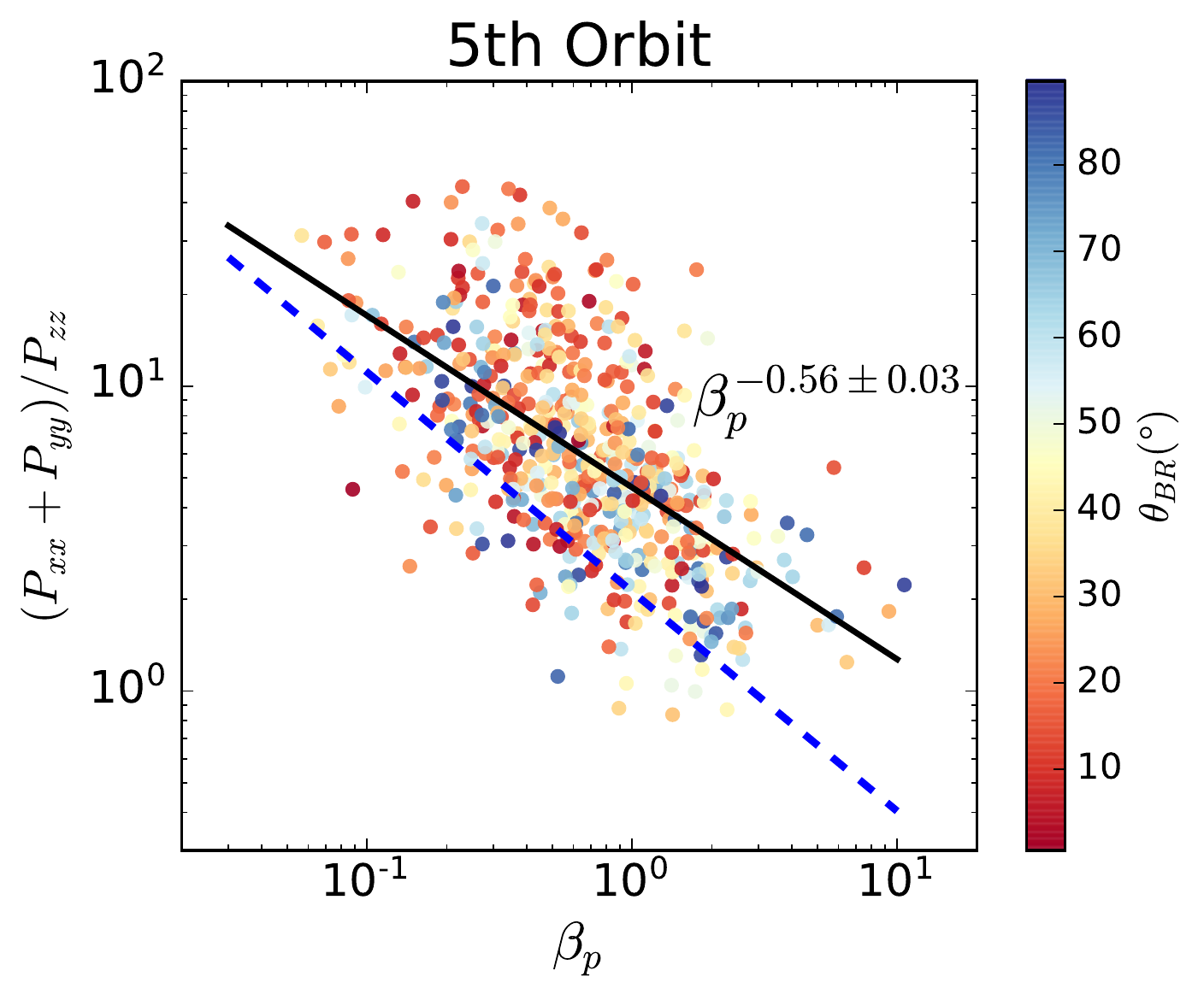}%
\includegraphics[width=0.5\linewidth]{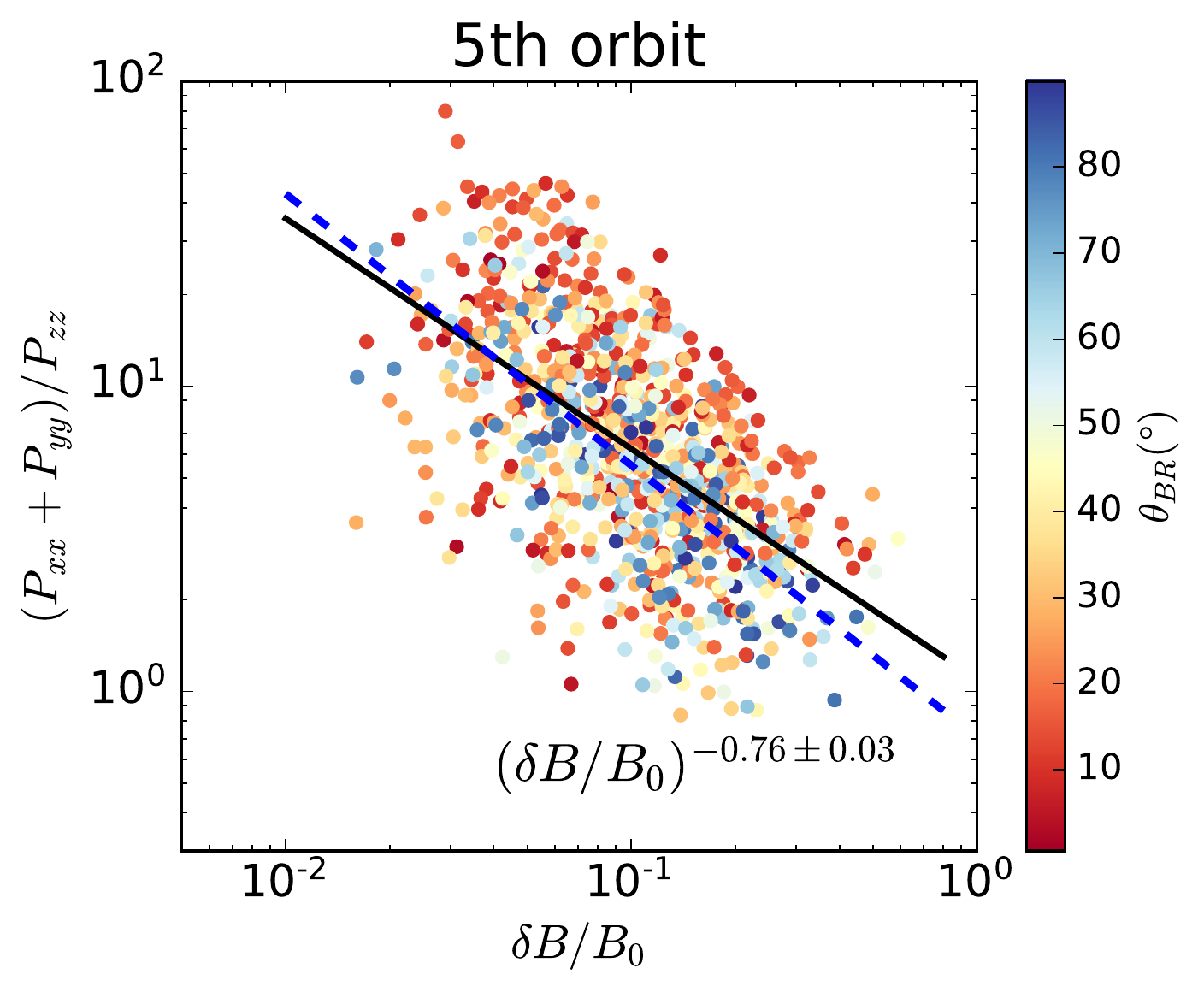}
	\caption{The transverse-to-longitudinal power anisotropy $(P_{xx}+P_{yy})/P_{zz}$ as functions of the proton plasma beta $\beta_p$ (left panel) and the ratio of magnetic fluctuation amplitude and mean magnetic field strength $\delta B/B_0$ (right panel) during \textit{PSP}'s fifth orbit. The color map indicates the $\theta_{BR}$ dependence for both panels. The two solid black lines denote power-law fits for the relation with $\beta_p$ and $\delta B/B_0$, respectively. The two blue dashed lines are from \cite{smith2006} and are used to fit the corresponding relation of intervals with open field lines at 1 AU.}\label{fig:view0404}
\end{figure}

For completeness, now consider the dependence of $(P_{xx}+P_{yy})/P_{zz}$ with the proton plasma beta $\beta_{p}$ and $\delta B/B_0$ (the ratio between magnetic fluctuation amplitude and mean field strength) following \cite{smith2006, pine2020solar}. The left panel of Figure \ref{fig:view0404} shows that the turbulence tends to be more compressive at higher plasma beta and the anisotropy follows a $4.65\beta_p^{-0.56}$ relation. The right panel shows that the compressibility is higher when the magnetic fluctuation amplitude is larger with a power-law relation $1.1(\delta B/B_0)^{-0.76}$. The $(P_{xx}+P_{yy})/P_{zz}$ vs.\ $\delta B / B_0$ relation shown in the right panel of Figure \ref{fig:view0404} is remarkably consistent with the results reported by Smith et al. (2006) using \textit{ACE} data (displayed as the blue dashed line in the figure). The $(P_{xx}+P_{yy})/P_{zz}$ vs.\ $\beta_p$ relation in Figure \ref{fig:view0404} has a power-law slope that is also roughly consistent with \cite{smith2006} though the multiplication factor is somewhat different. The angle $\theta_{BR}$ may play a role in explaining the difference as small-angle intervals tend to be much more common closer to the Sun and these intervals appear to deviate from the reference line (blue dashed line) more than large $\theta_{BR}$ intervals. We also note that the proton temperature used in this study is from the SPC instrument \citep{Kasper2016}, which does not measure the 3D proton temperature as does the SWEPAM instrument on \textit{ACE} \citep{mccomas1998}. Due to the temperature anisotropy, it is likely that the measured proton temperature is overestimated when the angle $\theta_{BR}$ is quasi-perpendicular, especially when $\beta_p$ is small \citep{huang2020proton}. As a result, the $\beta_p$ value may be overestimated in Figure \ref{fig:view0404}, which may also contribute to the difference from the \cite{smith2006} result. 
 
As noted by \cite{smith2006}, the two parameters $\beta_p$ and $\delta B / B_0$ are correlated themselves and $\beta_p$ seems to be the determining factor that establishes the transverse-to-longitudinal power anisotropy. The negative correlation between $(P_{xx}+P_{yy})/P_{zz}$ and beta is consistent with simulations such as those by \cite{matthaeus1996} but it is still not well understood theoretically.
There is a possibility that the parameter dependence of $(P_{xx}+P_{yy})/P_{zz}$ shown in Figure \ref{fig:view0404} could explain the left panel of Figure \ref{fig:psd1} as more small $\theta_{BR}$ intervals tend to be observed closer to the Sun where $\beta_p$ and $\delta B / B_0$ also tend to be smaller. However, we emphasize that it may also be possible that the compressibility $(P_{xx}+P_{yy})/P_{zz}$ depends on the wavevector directly as the NI theory suggests in certain limits \citep[e.g.,][]{Zank1993PF}.

To summarize, Figure \ref{fig:view1103} and \ref{fig:psd1} both suggest that the turbulence properties (spectral index and magnetic compressibility) depend on the field orientation $\theta_{BR}$. It is also possible that transport properties of the fluctuations play a role, but this angular dependence may obscure the true nature of turbulence. In other words, an apparent change in observed turbulence properties may be either due to parameters such as plasma beta and fluctuation amplitude as shown in Figure \ref{fig:view0404}, or the radial distance and sampling trajectory of the spacecraft. The latter is due to the wavevector anisotropy of turbulence \citep[e.g.,][]{matthaeus1990, horbury2012}. While the spectral index, magnetic compressibility, and wavevector anisotropy are typically regarded as independent issues, anisotropy may affect the spectral index [e.g., the two-component spectral theory of \cite{Zank2020ApJ} relates wavenumber anisotropy and spectral index, as does the critical balance theory of \cite{goldreich1995}], which includes identifying potential compressible magnetic fluctuations in the chosen geometry \citep{bieber1996, saur1999}. One needs to disentangle the effects of sampling direction and radial distance (transport effects) to study the intrinsic properties of turbulence.

\subsection{2D and slab model anisotropy}\label{sec:2Dslab}
\begin{figure}[htbp]
\centering
\includegraphics[width=0.5\linewidth]{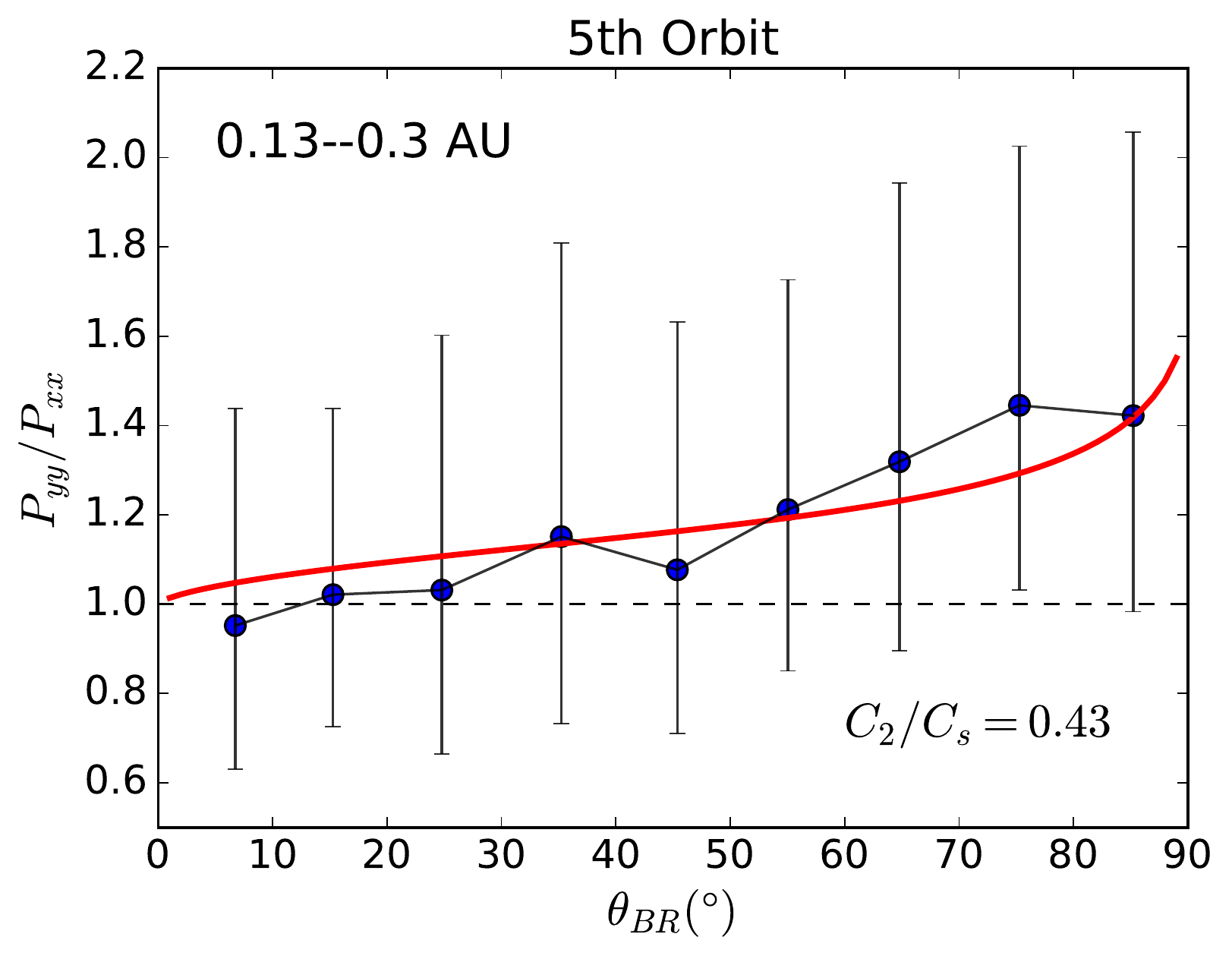}%
\includegraphics[width=0.5\linewidth]{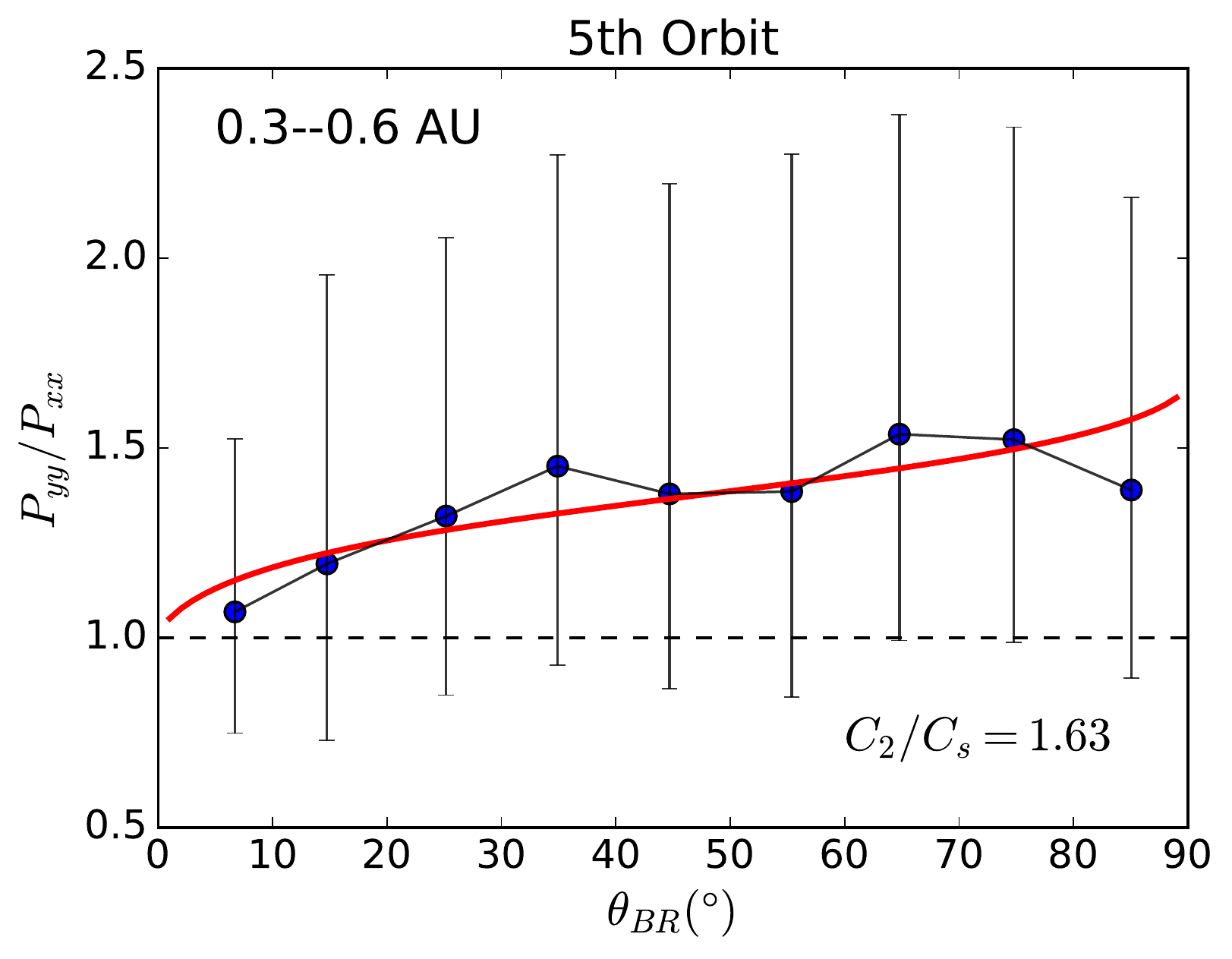}
\caption{Statistics of the transverse power anisotropy $P_{yy}/P_{xx}$ as a function of the sampling angle $\theta_{BR}$ during \textit{PSP}'s fifth orbit. The data are split into two groups by radial distance from the Sun. The left panel is for radial distances between 0.13 and 0.3 AU; the right panel is for radial distances between 0.3 and 0.6 AU. The red curves are the model results.}\label{fig:psd}
\end{figure}

One way to extract the wavevector anisotropy information is based on the 2D and slab turbulence model. Here, we present results based on the methodology discussed in Section \ref{sec:m} \citep{bieber1996}. Figure \ref{fig:psd} shows the average anisotropy $P_{yy}/P_{xx}$ as a function of $\theta_{BR}$. Here the averages and errors are calculated through logarithmic averaging due to the dynamic range of $P_{yy}/P_{xx}$ that can vary from 0.4 to 6 as shown in Figure \ref{fig:psd1}. By doing it this way, we can remove the influence of a few large values and make the average more representative of the overall dataset. Data from \textit{PSP}'s fifth orbit are presented here as examples and the data are split into two groups based on radial distance. The left panel of figure \ref{fig:psd} includes data taken between 0.13 and 0.3 AU while the right panel includes data between 0.3 and 0.6 AU. For both groups, the data intervals are binned in 9 bins that are 10$^\circ$ wide in each. As shown in the figure, the averaged $P_{yy}/P_{xx}$ increases with the increasing sampling angle $\theta_{BR}$ in both groups, which is consistent with the 2D and slab turbulence model. 

As discussed before, the exact relation between $P_{yy}/P_{xx}$ and $\theta_{BR}$ depends on the power ratio between 2D and slab components $C_2/C_s$. Fitting the model Equation \eqref{eq:Pyyxx} to the data yields the best-fit parameter $C_2/C_s$. The best-fit models are plotted as the red curves in both panels, which agree with the data reasonably well. We find the ratio to be $C_2/C_s = 0.43$ (or 30\%:70\%) between 0.13 and 0.3 AU and $C_2/C_s = 1.63$ (or 62\%:38\%) between 0.3 and 0.6 AU. At smaller radial distances, the slab component appears to be more important relative to the 2D component.

The analysis above is repeated for the other \textit{PSP} orbits and the results are compiled in Table \ref{tab:c2cs}.
In all orbits that we analyzed, we find a radial dependence of the wavevector anisotropy: the power fraction in the 2D components is larger further away from the Sun and the slab component is more dominant closer to the Sun. As shown earlier in Figure \ref{fig:view1103}, the spectral index $|q|$ tends to decrease at a smaller radial distance. We try to use a different spectral index ($q = -1.5$) in Equation \ref{eq:Pyyxx} for the small radial distance range ($\le$ 0.3 AU), which is also shown in Table \ref{tab:c2cs}. Changing the spectral index makes the $C_2/C_s$ ratio larger (or the fraction of 2D component is larger), but the conclusion that the $C_2/C_s$ ratio increases with distance still holds. At larger radial distances (0.3--0.6 AU), the 2D fluctuation power usually dominates containing $\sim 60\%$ of the total transverse fluctuation power, which is slightly less than the results obtained by \cite{bieber1996} with $\sim 80\%$ 2D contribution. An exception is the 7th orbit where the 2D fluctuation power is only 40\%. The exact reason is unclear, but could be due to mixing of fast solar wind or different origins of slow solar wind. Since the solar wind is out of the field of view of SPC during most of the time period of the 7th orbit, we do not have enough plasma data for further study. 
The region within 0.3 AU from the Sun has not been explored by previous observations. Our results indicate that the slab component dominates in this new regime, accounting for about 60\%--80\% of the total transverse power.

\begin{deluxetable*}{cccc}
\tablecaption{Wavevector anisotropy in different \textit{PSP} orbits \label{tab:c2cs}}
\tablehead{\colhead{\qquad Orbit \qquad} & \colhead{\qquad $C_2/C_s$ ($\le$ 0.3 AU) \qquad} & \colhead{\qquad $C_2/C_s$ ($\le$ 0.3 AU) \qquad} & \colhead{\qquad $C_2/C_s$ (0.3--0.6 AU) \qquad}\\  & $q=-5/3$ & $q=-3/2$ & $q=-5/3$ } 
  \startdata
  \# 1  &  0.47 (32\% : 68\%) &  0.75 (43\% : 57\%)  &  1.70 (63\% : 37\%)\\
  \# 2  &  0.11 (10\% : 90\%) &  0.20 (17\% : 83\%)  &  1.63 (62\% : 38\%)\\
  \# 4  &  0.28 (22\% : 78\%) &  0.47 (32\% : 68\%)  &  1.56 (61\% : 39\%)\\
  \# 5  &  0.43 (30\% : 70\%) &  0.72 (42\% : 58\%)  &  1.63 (62\% : 38\%)\\
  \# 6  &  0.40 (29\% : 71\%) &  0.52 (34\% : 66\%)  &  1.50 (60\% : 40\%)\\
  \# 7  &  0.25 (20\% : 80\%) &  0.49 (33\% : 57\%)  &  0.67 (40\% : 60\%)\\
  \enddata
  \tablecomments{The perihelion distance to the Sun is different for different \textit{PSP} orbits. The exact radial distance range in the second column also varies with orbit, which we list as follows: 0.17--0.3 AU (Orbits 1 and 2); 0.13--0.3 AU (Orbits 4 and 5); 0.10--0.3 AU (Orbits 6 and 7).}
\end{deluxetable*}

Based on the 2D-slab ratio that we obtain, it is likely that the total transverse power $P_{xx}+P_{yy}$ increases with $\theta_{BR}$ (at least at $R >0.3$ AU where $C_s$ does not dominate $C_2$). This allows us to draw further conclusions about the nature of the compressible component of turbulence from the left panel of Figure \ref{fig:psd1}. As the ratio $(P_{xx}+P_{yy})/P_{zz}$ decreases with angle, it means that $P_{zz}$ should increase with $\theta_{BR}$ even faster than the transverse power $P_{xx}+P_{yy}$, suggesting that the compressible component $P_{zz}$ may also contain a significant 2D component. Such 2D-like compressible fluctuations are allowed in the NI theory in the $\beta \ll 1$ limit \citep{Zank1993PF}. A more quantitative analysis of the compressible component is deferred to a future work.

\section{Conclusions}

The \textit{Parker Solar Probe} provides unprecedented opportunities to study the evolution of turbulence in the inner heliosphere. However, the presence of the turbulence wavevector anisotropy complicates the interpretation of observations.
In this paper, we present an analysis of MHD-scale magnetic fluctuations observed by \textit{PSP}. We demonstrate that observed turbulence properties such as the spectral index and the magnetic compressibility depend on the sampling angle of the spacecraft due to wavevector anisotropy. This angular dependence may obscure how turbulence is intrinsically related to parameters, such as, radial distance, plasma beta, fluctuation amplitude, etc.
To unravel the effects of the sampling angle, we analyze the wavevector anisotropy in the context of the 2D and slab turbulence model. The power ratio between 2D and slab components is determined by the data and a radial dependence is discovered.

The main conclusions of the paper are listed as follows.
\begin{enumerate}
\item The observed magnetic power spectrum is shallower closer to the Sun. The radial dependence may be related to the sampling angle $\theta_{BR}$ (between the mean magnetic field and the radial direction) dependence because the sampling angle tends to decrease close to the Sun due to the form of the Parker spiral magnetic field.
\item The magnetic compressibility tends to become weaker closer to the Sun, and is positively correlate with the plasma beta and the magnetic fluctuation amplitude, consistent with previous observations \citep[e.g.,][]{smith2006, chen2020}. However, these dependencies can evolve with the sampling angle dependence.
\item The spectral anisotropy of the transverse fluctuations is generally consistent with the 2D plus slab turbulence model.
\item We determine the fraction of power in the 2D versus slab component using the method proposed by \cite{bieber1996}. Exact values for each orbit are listed in Table \ref{tab:c2cs}.
\item The fraction of power in 2D fluctuations is smaller closer to the Sun during all \textit{PSP} orbits that we analyzed, based on the \cite{bieber1996, saur1999} approach.
\end{enumerate}

Finally, we list some caveats for the analysis presented in this study, which should be addressed in subsequent studies. 

1) Our analysis of the 2D and slab turbulence decomposition assumes a single power-law index that does not depend on the angle $\theta_{BR}$, i.e., $q$ in Equation \eqref{eq:Pyyxx} is a constant. While this assumption simplifies the analysis, there is theoretical and empirical evidence suggesting that the spectral index may depend on the sampling angle \citep[e.g.,][and also our Figure \ref{fig:view1103}]{horbury2008}. For future studies, our analysis should incorporate the possibility of a spectral index anisotropy \citep{Zank2020ApJ}.

2) We did not distinguish fluctuations according to the origin of the solar wind, i.e., open versus closed field regions, and this may reflect the different turbulence mechanisms thought to heat the solar corona and subsequently accelerate the solar wind. For example, \cite{zank2018theory} suggest that it is primarily the dissipation of 2D fluctuations that heat the corona, which would make derived $C_2/C_s$ ratio observed within $\sim$0.3 AU understandable since it is mainly 2D fluctuations that are absorbed. Under the mechanism proposed by \cite{matthaeus1999}, i.e., the dissipation of turbulence generated by counter-propagating slab fluctuations, one might expect the opposite result \citep[][]{zank2021turbulence}.

3) The third caveat is the neglect of the radial dependence within the range $R \le$ 0.3 AU and $0.3 < R < 0.6$ AU \citep{adhikari2021a}. This caveat is difficult to overcome. If we refine the binning in $R$, the data in each radial distance bin may not cover the entire angular range, leading to more ambiguity in fitting the model.


4) We did not consider the different plasma beta regimes when applying the \cite{bieber1996} analysis. From a theoretical perspective, the underlying character of MHD turbulence is fundamentally different in a $\beta \gg 1$ plasma than a $\beta \sim 1$ or $\ll$ 1 plasma. The 2D-slab decomposition does not emerge in the former case where the governing incompressible MHD equations are fully 3D.

5) Since solar wind velocity data are not always available, we use $\theta_{BR}$ to represent the angle between the wavevector and the mean magnetic field. Although the solar wind velocity is typically dominated by the radial component, the nonradial components can have a significant contribution close to the Sun \citep{kasper2019}, making the difference between $\theta_{BR}$ and $\theta_{VB}$ (angle between solar wind speed in the spacecraft frame and the mean magnetic field) slightly larger. Thus, at this point the results may not be very accurate near the perihelion.

The five caveats discussed above make further detailed analysis of this problem necessary. Nonetheless, our method shows for the first time clear wavevector anisotropy in the new regime being explored by \textit{PSP}.
The anisotropy of solar wind turbulence provides considerable insight into its nature. Our work shows that the solar wind data measured by PSP is consistent with the 2D and slab model as suggested by the nearly incompressible MHD theory. For the first time, we have identified the changing ratio between 2D and slab fluctuations in the inner heliosphere. Apparently, a recent work by \cite{Bandy2021} also uses the \cite{bieber1996} approach to analyze the 2D-slab ratio using \textit{PSP} data. Their results are qualitatively consistent with ours although they do not distinguish between different radial distance regions. \cite{Bandy2021} construct the 2D correlation function and find a shorter parallel correlation length than perpendicular correlation length, which is opposite to typical observations for slow solar wind at 1 AU \citep[e.g.,][]{matthaeus1990, dasso2005}. This is also consistent with our conclusion of an increasing slab fraction closer to the Sun. Our results will further the understanding of how turbulence is generated and transported in the solar wind, and will guide the development of future solar wind turbulence models.

\section*{\leftline{Acknowledgement}}
\begin{acknowledgments}
We acknowledge the partial support of the NSF EPSCoR RII-Track-1 Cooperative Agreement OIA-1655280, a NASA award 80NSSC20K1783 and a NASA Parker Solar Probe contract SV4-84017. We thank the NASA Parker Solar Probe SWEAP team led by J. Kasper and FIELDS team led by S. D. Bale for use of data. 	
\end{acknowledgments}

\bibliography{turbu-aniso}{}
\bibliographystyle{aasjournal}



\end{document}